\documentclass[prb,reprint]{revtex4-1} 

\usepackage{amsmath}
\usepackage{amsfonts}
\usepackage{graphicx}
\usepackage{subfig}
\usepackage{physics}

\usepackage[outdir=./]{epstopdf}

\begin{document}

\title{A Flexible Positron Spectrometer for the Undergraduate Laboratory}

\author{Jason Engbrecht}
\email{engbrech@stolaf.edu}

\author{Nathaniel Hillson}
\affiliation{Department of Physics, St.\@ Olaf College, Northfield, MN 55057}

\date{\today}

\begin{abstract}
Positron physics touches on a wide-ranging variety of fields from materials science to medical imaging to high energy physics.  In this paper we present the development of a flexible positron annihilation spectrometer appropriate for the undergraduate laboratory.  Four NaI gamma-ray ($\gamma$-ray) detectors are connected to an oscilloscope-based data acquisition system.  Coupled with the software we developed, these detectors allow students to explore a variety of positron and $\gamma$-ray phenomena.  These include $\gamma$-ray energy spectroscopy, Compton scattering, PET scanning fundamentals, speed of light measurements with $\gamma$-rays, historically important polarimetry of annihilation radiation, 3-$\gamma$ annihilation radiation observations, and positron lifetime spectroscopy of materials.  We present the developed apparatus and examples of experiments it can perform here.
\end{abstract}

\maketitle

\section{Introduction}

With its discovery in 1932 by Carl Anderson, the positron began the era of antimatter in physics.  Unlike its antimatter counterparts, the antiproton and antineutron, the positron can be produced from radioactive sources.  As such positrons are available for a large array of applications.  PET scans,\cite{1} materials science,\cite{2,3} QED tests,\cite{4} fundamental symmetry experiments,\cite{4} and antihydrogen spectroscopy\cite{5,6,7,8} are some of the many areas in which positrons have found applications both historically and in the modern day.

The accessibility of positrons has also brought them to the teaching laboratory.  Experiments for undergraduate laboratory classes have been presented in the literature in the past.\cite{9,10,11,12,13,14,15} Each of these experiments in literature, while fine examples, focused on one aspect of positron physics with a dedicated apparatus.  

With this work we set out to build a single apparatus along with data acquisition software that would allow students to explore the variety of physics accessible with positrons.  With that in mind, we focus on detectors and acquisition software that provide maximum flexibility.  This paper presents the apparatus and software developed for this purpose.  Additionally, it presents examples of experiments performed with this apparatus.

\begin{figure}[h!]
\centering
\includegraphics[width=3.4in]{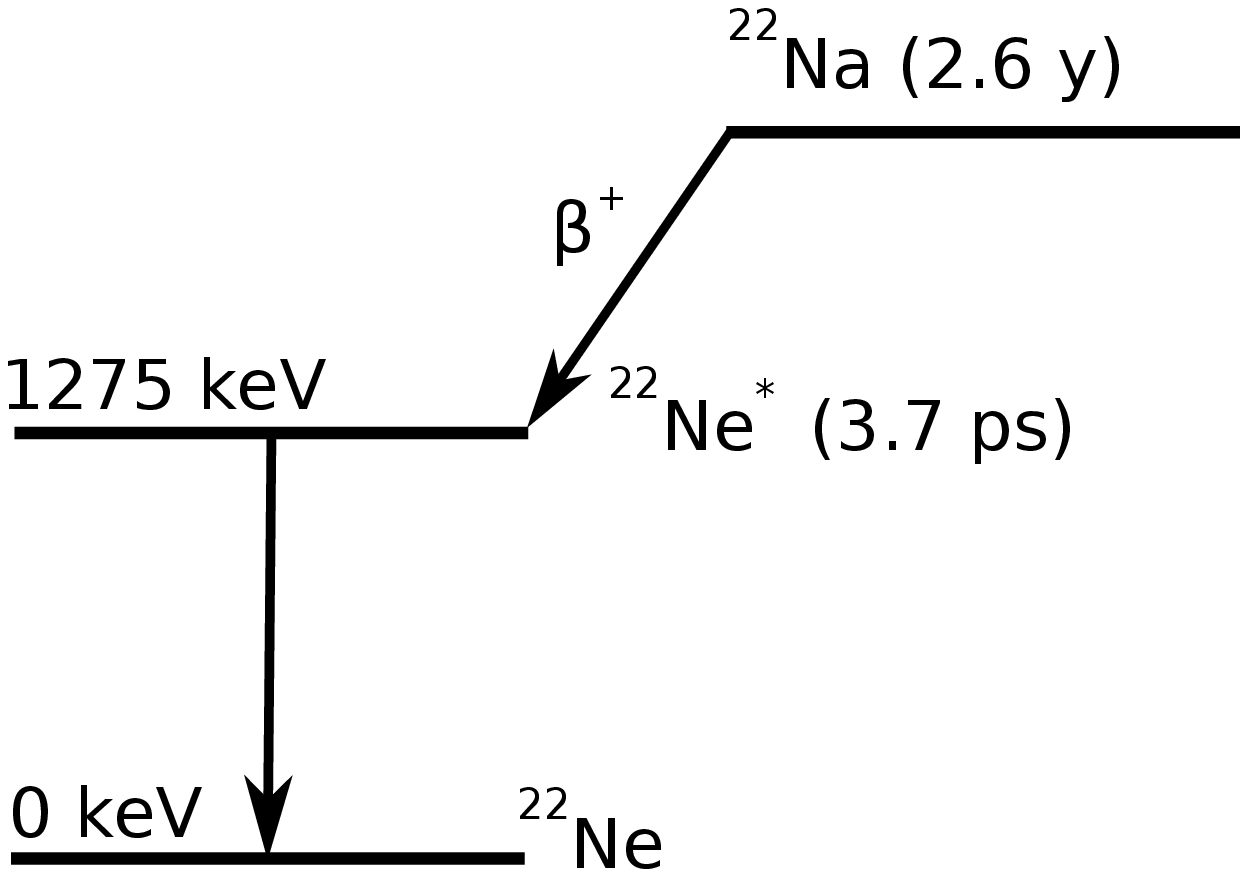}
\caption{Decay Scheme of \textsuperscript{22}Na.}
\label{nadecay}
\end{figure}

The existence of commercially available radioactive sources that produce positrons is key to making experiments such as those presented here accessible to the undergraduate laboratory.  The most common choice of source is \textsuperscript{22}Na with a decay scheme shown in Fig.~\ref{nadecay}. With a 2.6-year half-life, \textsuperscript{22}Na is long-lived enough that data rates are stable throughout the course of an undergraduate lab experience.  An additional property of value for \textsuperscript{22}Na is the production of a 1275 keV $\gamma$-ray shortly after the emission of the positron.  This $\gamma$-ray comes from the excited state of \textsuperscript{22}Ne that is formed during the emission of the positron.  The short lifetime of this excited state (3.7 ps) is essentially instantaneous for the purpose of all the experiments presented here. Thus, the detection of this 1275 keV $\gamma$-ray can serve as a positron creation signal in timing experiments.

The positron, once emitted, will do one of two things.  It can either find an electron and directly annihilate with it, or it can find an electron and bind with it, forming the electron-positron bound system called positronium (Ps).  Ps can form in either the S=0 state (para-Ps) or the S=1 state (ortho-Ps).  Direct annihilation and para-Ps annihilation both decay into 2 $\gamma$-rays.  Time scales for these 2-$\gamma$ annihilations are typically a few hundred ps.  

Ortho-Ps has a much longer lifetime of 142 ns in a vacuum because its spin does not allow it to decay into 2 $\gamma$-rays and instead it produces 3 $\gamma$-rays during annihilation.  However, in the dense medium of a solid, the positron in the electron-positron pair will usually find an alternate electron that is more favorably aligned to allow it to decay via a 2-$\gamma$ annihilation.  Thus, in most solid materials any ortho-Ps formed will have its lifetime reduced, typically to a few ns.

The net result of all these positron interactions is that the overwhelming majority of positrons will annihilate with an electron within a few ns and form 2 $\gamma$-rays.  Conservation of energy and momentum demand that these 2 $\gamma$-rays will be of equal energy (511 keV) and will travel in anti-parallel directions.  This annihilation pair has some distinct experimental advantages when compared to other $\gamma$ radiation from radioactive sources.  Detecting both $\gamma$-rays in the pair can be used to locate the source of the radiation, to reduce random background radiation, and to allow for polarimetry.  All of these advantages are demonstrated in experiments presented in this paper.

\section{Apparatus}

In order to facilitate wider adoption of the presented work, when designing the apparatus for laboratory attention was paid to cost and use of commercially available parts.  It is estimated that the entire experiment can be reproduced for approximately \$11,000.  A moderate amount of 3D printing and machining was used to create the experimental setup using skills and tools available in most department machine shops.  The custom-built power supply used in this work can be straightforwardly reproduced or replaced with commercial bench top supplies for a reasonable cost increase. Table~\ref{equipment} shows the equipment list for the present work as well as estimated costs.

\begin{table}[h!]
\centering
\caption{Equipment List and Costs}
\begin{ruledtabular}
\begin{tabular}{l r}
Item & Cost\\
\hline
NaI Detectors & \$5,500 \\
USB Oscilloscope and Cables & \$3,500 \\
\textsuperscript{22}Na Sources & \$500 \\
Lead Collimator & \$200 \\
HV Power Supply & \$1,000 \\
Miscellaneous Hardware and 3D Printing & \$200 \\
\hline
Total & \$10,900 \\
\end{tabular}
\end{ruledtabular}
\label{equipment}
\end{table}

\begin{figure}[h!]
\centering
\includegraphics[width=3.4in]{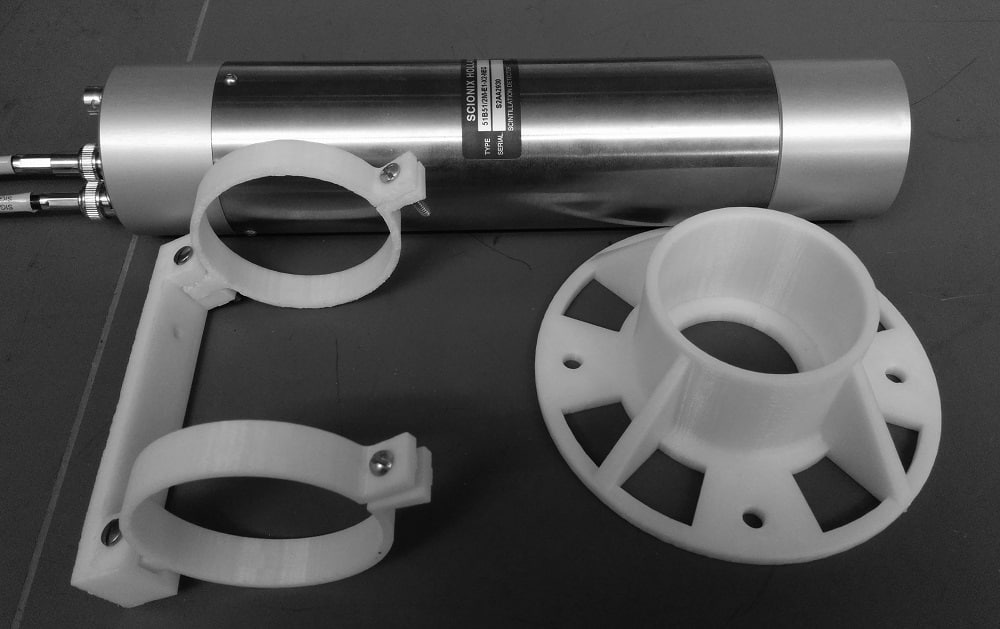}
\caption{NaI detector with integrated photomultiplier shown with two 3D printed holders.}
\label{detectorandholders}
\end{figure}

The foundation of the experimental apparatus is four 51 mm by 51 mm cylindrical NaI scintillation crystals paired with photomultipliers.  We used NaI-based detectors to give the best balance of energy resolution, timing resolution, detection efficiency and affordability for $\gamma$-ray detection.   Modern crystals such as BaF$_2$ or CeBr$_3$ have superior timing properties to NaI. However, pricing of these crystals was between 3 and 10 times higher than NaI so we decided to pursue the more economical model.   The NaI detectors can be reconfigured into a variety of orientations for different experiments.  To facilitate this, a number of 3D printed mounts were produced to secure and protect the detectors. A detector and two of these holders are shown in Fig.~\ref{detectorandholders}.  A custom-made 4-channel -2 kV variable power supply was built to power the detectors.   After pulse processing, these detectors exhibited energy resolution of  8.4\% FWHM (at 511 keV) and a timing resolution of 2.4 ns.

The electrical pulses from the NaI detectors are traditionally processed by dedicated pulse-processing electronics.  As an example, timing between two detectors is often done using two Constant-Fraction Single Channel Analyzers that locate the pulses in time.  These are fed to a Time-to-Analog Converter to produce a new pulse proportional in height to the time between the signals.  Finally, this is sent to a Multi-Channel Analyzer that digitizes the pulse for a computer.  This electronics system alone can easily cost about \$10,000.  In order to achieve the flexibility desired for this apparatus, an alternative needed to be pursued.

To reduce costs and substitute for the traditional pulse processing electronics, we chose to use a USB-based oscilloscope.  The scope chosen was the 4-channel PicoScope 6402D, a 250 MHz bandwidth scope with a sample rate of 5 GS/sec. The sample rate is shared between active channels so that with all channels active the effective sample rate is 1.25 GS/sec. This scope allows the computer to capture the raw pulse data for each detection event and process it in software for energy and timing information.  The scope has sufficient capabilities such that, when paired with NaI detectors, software can analyze pulses with the same energy and timing precision of dedicated electronics.  The major sacrifice is data throughput.  Our system currently achieves throughputs of approximately 1000 events/sec compared to dedicated systems that can achieve 10--100 times faster throughput.  However, our achieved throughput is sufficient for all the experiments proposed here.

We utilize two types of commercially available \textsuperscript{22}Na radioactive sources to produce positrons.  For most experiments, only the $\gamma$-rays from annihilation are needed. For these, a 10 $\mu$Ci plastic source in which the positron primarily annihilates in the source encapsulating material is used.  For experiments in which the positrons need to exit the source, a 2 $\mu$Ci source in which the \textsuperscript{22}Na is sealed under a thin sheet of Mylar is used.   With a 2.6 year half-life, these sources will need to be replaced about every 3 years.

\begin{figure}[h!]%
    \centering
    \subfloat[Full View]{{\includegraphics[width=1.6in]{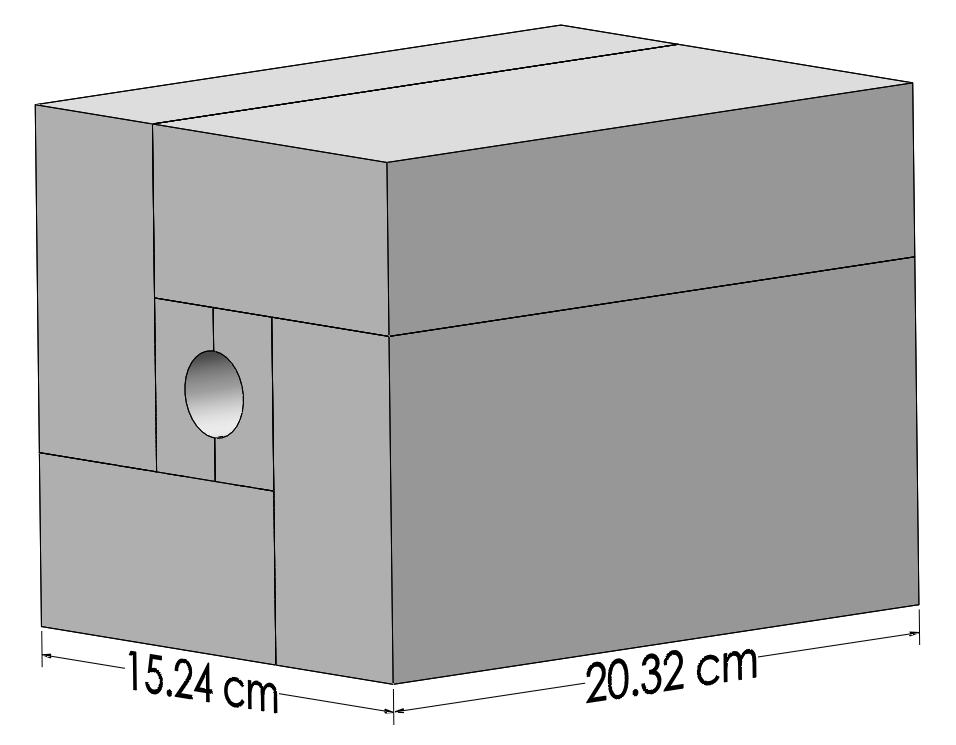} }}%
    \qquad%
    \subfloat[Cross Sectional View]{{\includegraphics[width=1.6in]{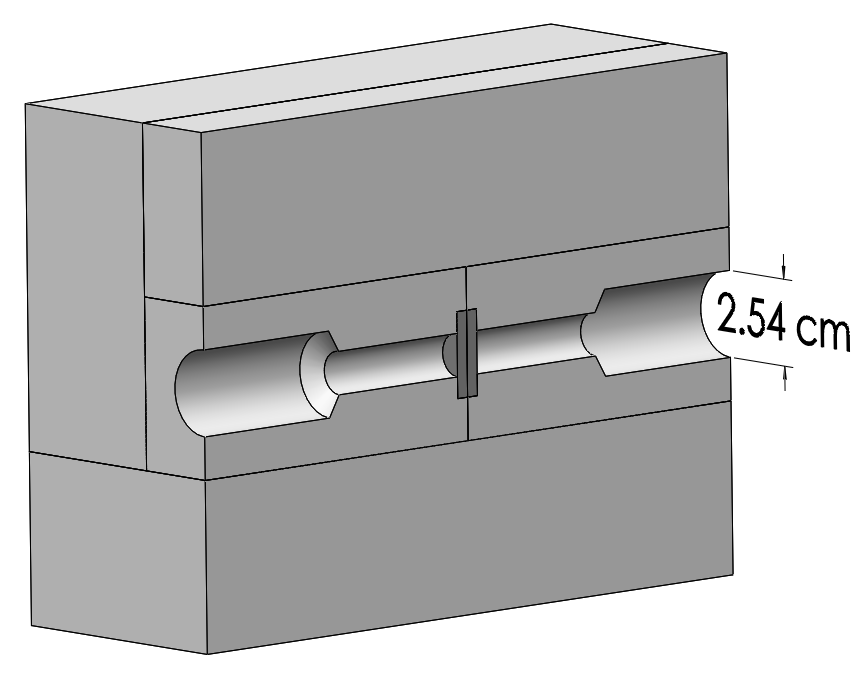} }}%
    \caption{Lead house used to produce collimated back-to-back $\gamma$-rays from 2-$\gamma$ annihilation radiation.}%
    \label{leadtelescope}%
\end{figure}

For a number of experiments, the annihilation radiation needs to be collimated.  For these experiments we have a lead house for $\gamma$-ray collimation in which two 10 $\mu$Ci sources are placed in the center, as shown in Fig.~\ref{leadtelescope}.

\section{Software}

The software for these experiments was developed using the LabView development  platform.  In developing the software we wanted to avoid experiment-specific interfaces that might inhibit the students' opportunity to explore.  Instead, we developed a set of general spectroscopic tools that could be applied to a variety of experiments.

\begin{figure}[h!]%
    \centering
    \subfloat[Original Pulse and Inverted/Timeshifted Pulse]{{\includegraphics[width=1.6in]{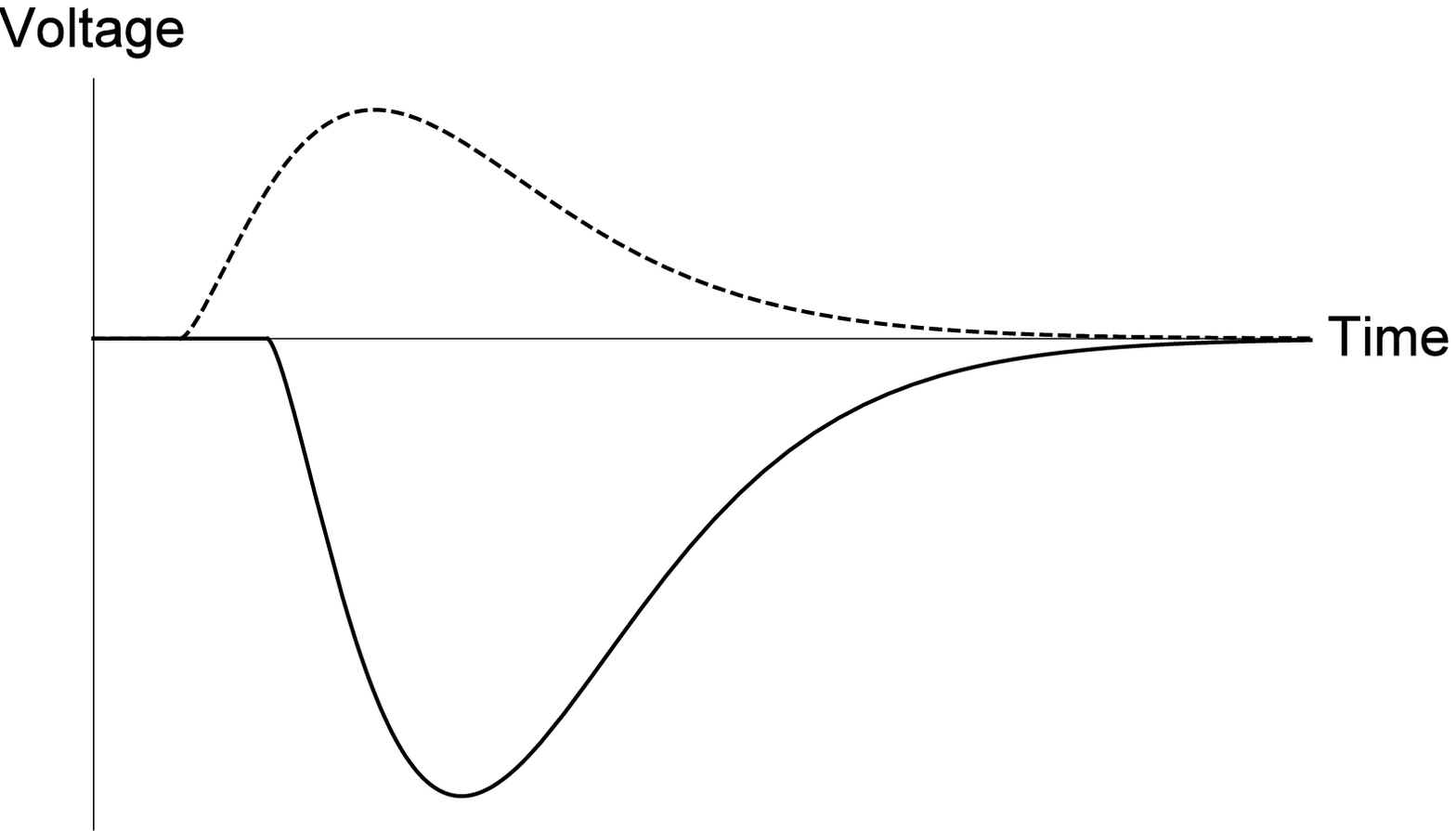} }}%
    \qquad%
    \subfloat[Added Pulses]{{\includegraphics[width=1.6in]{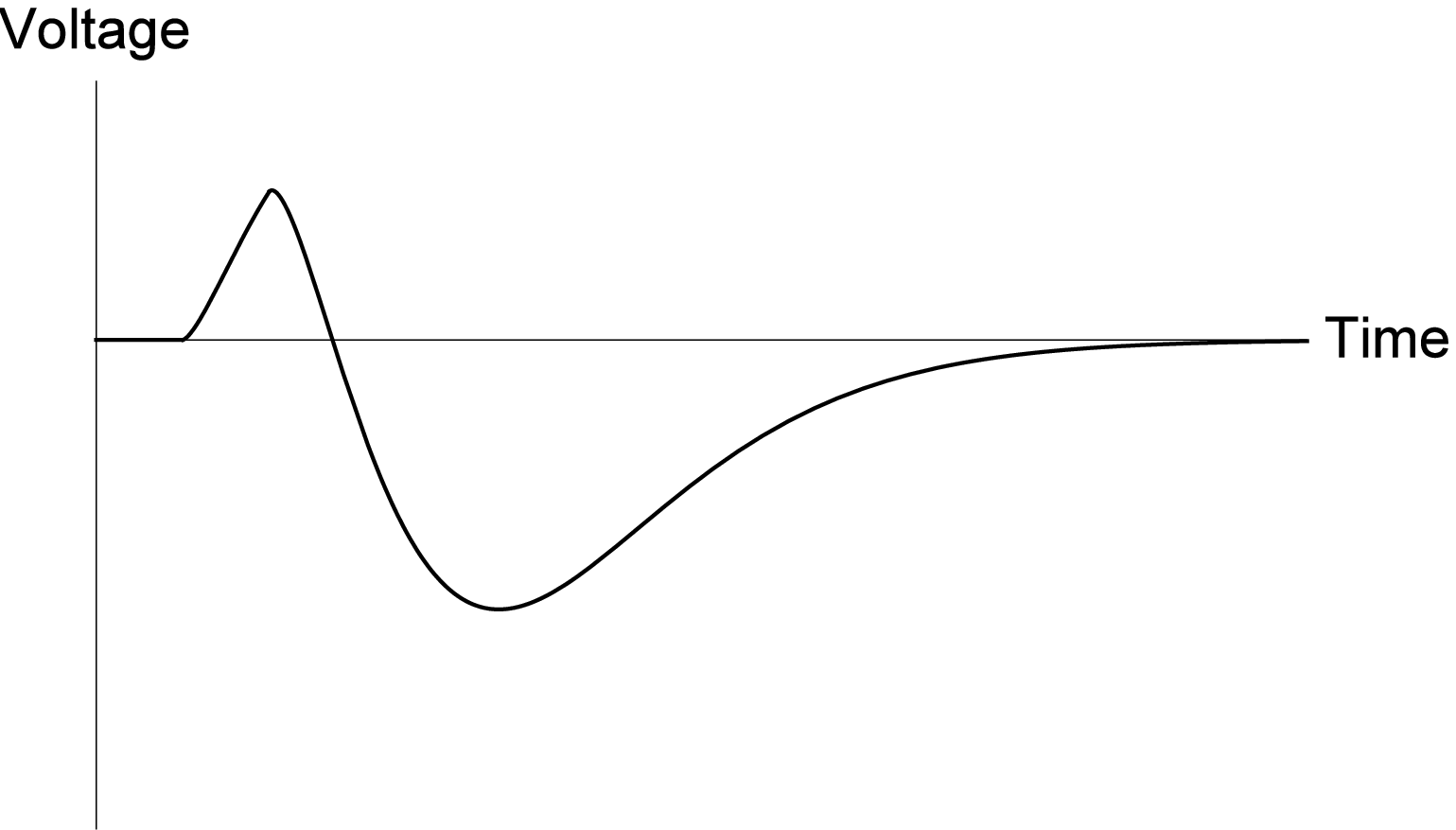} }}%
    \qquad%
    \subfloat[Integrated Pulse]{{\includegraphics[width=1.6in]{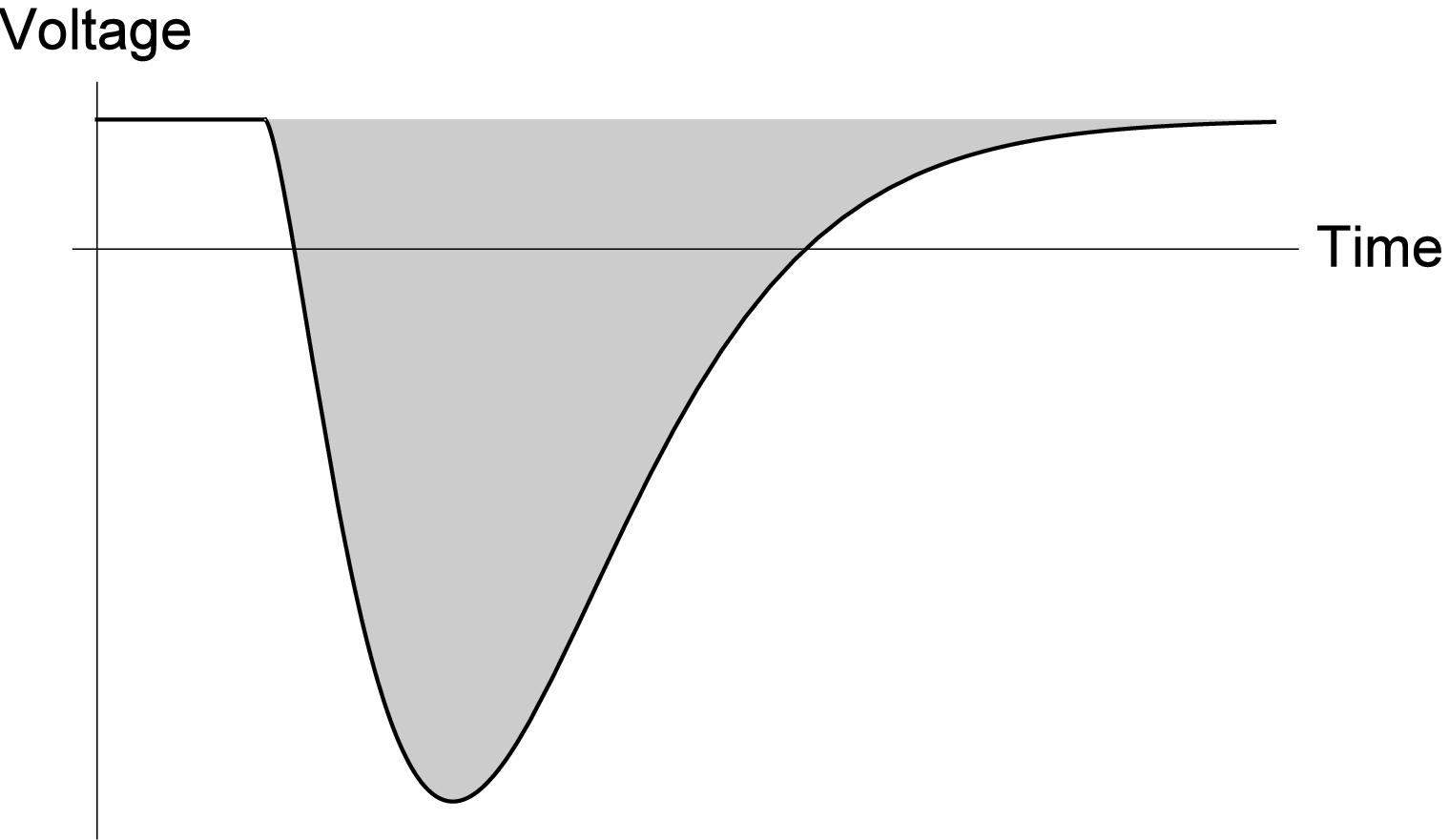} }}%
    \caption{(a) The process of Constant Fraction Discrimination in which a pulse is multiplied by a fraction and time shifted. (b) The result of adding the two signals in (a) creating a zero crossing that can be used for timing. (c) The shaded area represents the integration of a pulse while correcting for small zero offsets. The result is proportional to the energy deposited by the detected $\gamma$-ray.}%
    \label{pulseprocessing}%
\end{figure}

As data is acquired from the oscilloscope, the software performs analysis of the pulses from each detector.  Figure~\ref{pulseprocessing} shows how this process works.  Pulse area is proportional to the energy of the detected $\gamma$-ray and thus a simple numerical integration with background correction is used to determine the $\gamma$-ray energy. In order to determine the timing of each pulse a constant fraction method is used. In this method the pulse is inverted, scaled and time shifted before being added to the original pulse.  The result is then examined for a zero crossing to locate the pulse in time.  Once each pulse is analyzed for its timing and energy information, the original oscilloscope trace is erased from memory.  The energy and timing information are cataloged for every pulse to be used in later analysis.

\begin{figure}[h!]
\centering
\includegraphics[width=3.4in]{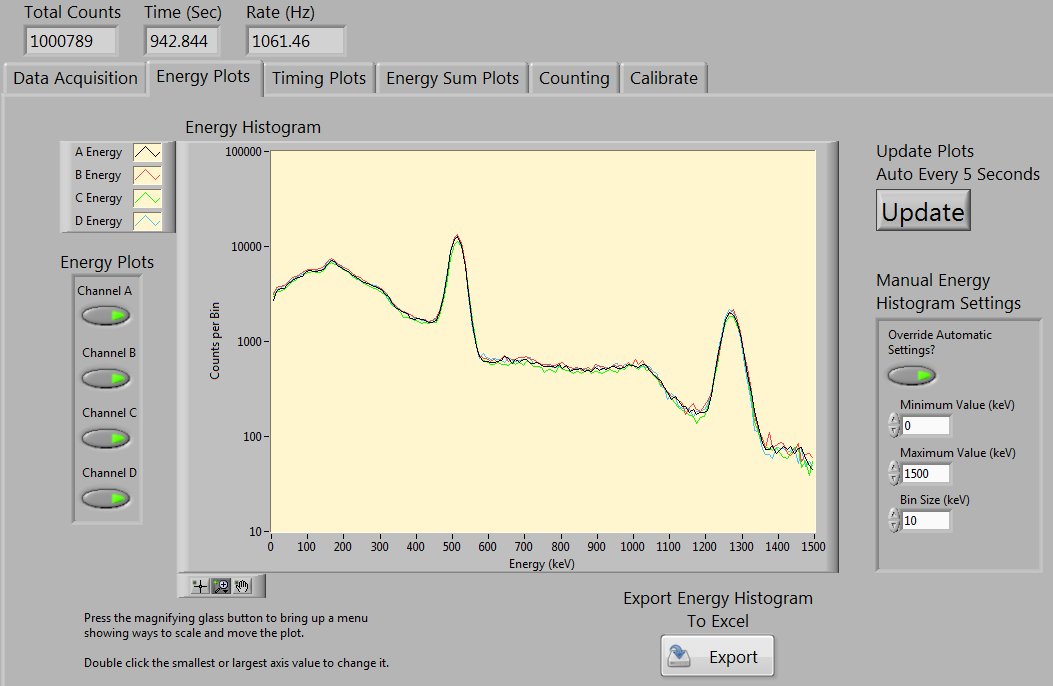}
\caption{The user interface for the data acquisition software showing the Energy Plots panel.}
\label{energypanel}
\end{figure}

The user interface consists of 6 selectable panels that provide interfaces to users for examining the pulse energy and timing information stored in memory.  Figure \ref{energypanel} shows the program with the Energy Plots panel chosen. 

The panels labeled Data Acquisition and Calibrate are used to prepare the acquisition system for acquiring data. The Data Acquisition panel allows the user to choose how data is acquired.  Data can be acquired until user intervention or until presets for acquisition time or event numbers.  From the panel the user can determine which detectors are being used and any coincidences that are required.  In this panel, all available event data (timing and energy) can be downloaded for offline analysis.  The Calibration panel allows the user to calibrate the area/energy conversion needed for turning the pulse areas into energy measurements.  This is an automated process completed by simply putting a \textsuperscript{22}Na source near all of the detectors and using the software to activate the calibration procedure.

The Energy Plots, Timing Plots, and Energy Sum Plots panels are all similar in their function.  They allow users to look at a variety of histograms of the data.   The Energy Plots and Energy Sum Plots both provide energy histograms but the Energy Sum Plots panel sums the energy of selected detectors before creating the histogram of the results.  The Timing Plots panel creates a histogram of the time between events in two different detectors.  This timing histogram can also be filtered by energy to allow only certain types of events to be included. The users have control over the limits of the histograms and the binning of the histograms if desired.  The histograms' data can also be downloaded for offline use.

The Counting panel allows the user to count events that meet specified timing and energy restrictions.  Multiple settings can be examined simultaneously.

\section{Experiments}

The apparatus developed is currently planned to be placed into two laboratory courses at St.\@ Olaf College.  The first lab is our Modern Physics Laboratory for sophomore-level physics majors.  In this lab, students have two 3-hour sessions.   The second lab is our Advanced Laboratory for junior-level physics majors.  In this lab, students have three 3-hour sessions.  Each lab course has a unique set of experiments.   In the junior level laboratory there is also an option for students to pursue some of the experiments presented here in more depth for an additional 3 weeks.

\section{Modern Lab Experiments}

The goal of the Modern Lab experiments is to introduce students to the basics of the apparatus and software, as well as the fundamental properties of positron annihilation.  They are by design shorter experiments that can be completed in a straightforward manner.

\subsection{$\gamma$-Ray Spectroscopy of Annihilation Radiation}

\begin{figure}[h!]
\centering
\includegraphics[width=3.4in]{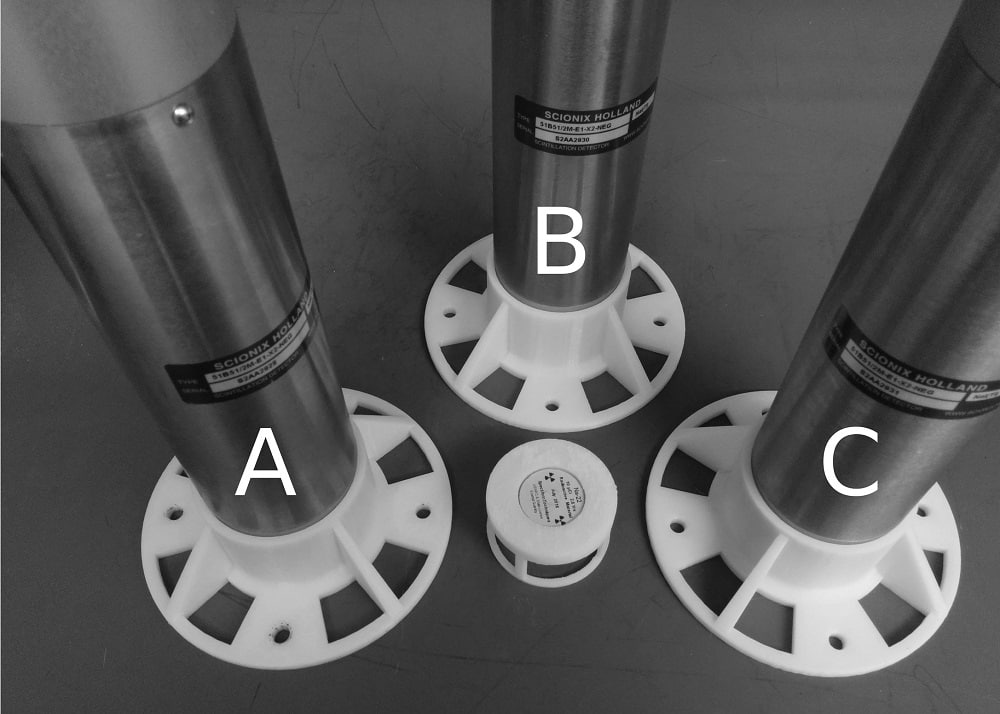}
\caption{Detector and source setup for $\gamma$-ray spectroscopy experiment.}
\label{gammaspectroscopysetup}
\end{figure}

This experiment introduces the students to the idea of $\gamma$-ray energy spectroscopy, in particular for annihilation radiation.  A \textsuperscript{22}Na source is placed in a holder on the table as a source of 1275 keV and 511 keV-pair $\gamma$-rays.  Two NaI detectors are placed in holders that allow for the students to vary the geometry of the system. In the software students can examine the effect of demanding coincidences between the two detectors. 

\begin{figure}[h!]
\centering
\includegraphics[width=3.4in]{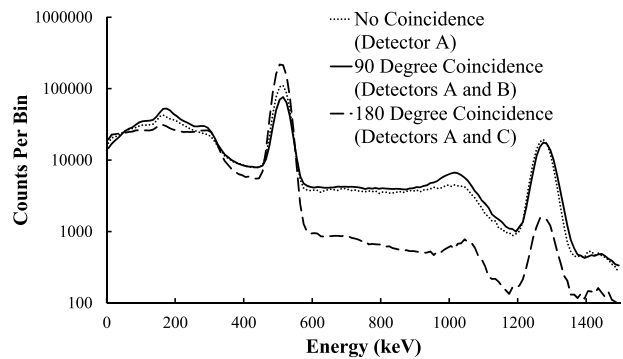}
\caption{$\gamma$-ray spectroscopy result using the detector setup of Fig.~\ref{gammaspectroscopysetup}.  Changing coincidence configurations alters the system's response to 511 keV vs 1275 keV $\gamma$-rays.}
\label{coincidenceplots}
\end{figure}

Figure~\ref{gammaspectroscopysetup} shows three detectors surrounding a \textsuperscript{22}Na source.  Energy spectra are acquired with a variety of detector coincidence configurations with the example results shown in Fig.~\ref{coincidenceplots}.  Using only one detector (detector A) compared to demanding a coincidence between detectors A and B shows a slight suppression of 511 keV $\gamma$-ray detections.  In contrast, with detectors A and C in coincidence, the 511 keV detections are greatly enhanced while the 1275 keV detections are suppressed.  These changes in the spectrum are due to the correlation in direction between the 511 keV $\gamma$-rays and the lack of such a correlation with the 1275 keV $\gamma$-rays.  Students are asked to consider and comment on these changes.

\subsection{PET Scanning Fundamentals}

\begin{figure}[h!]
\centering
\includegraphics[width=3.4in]{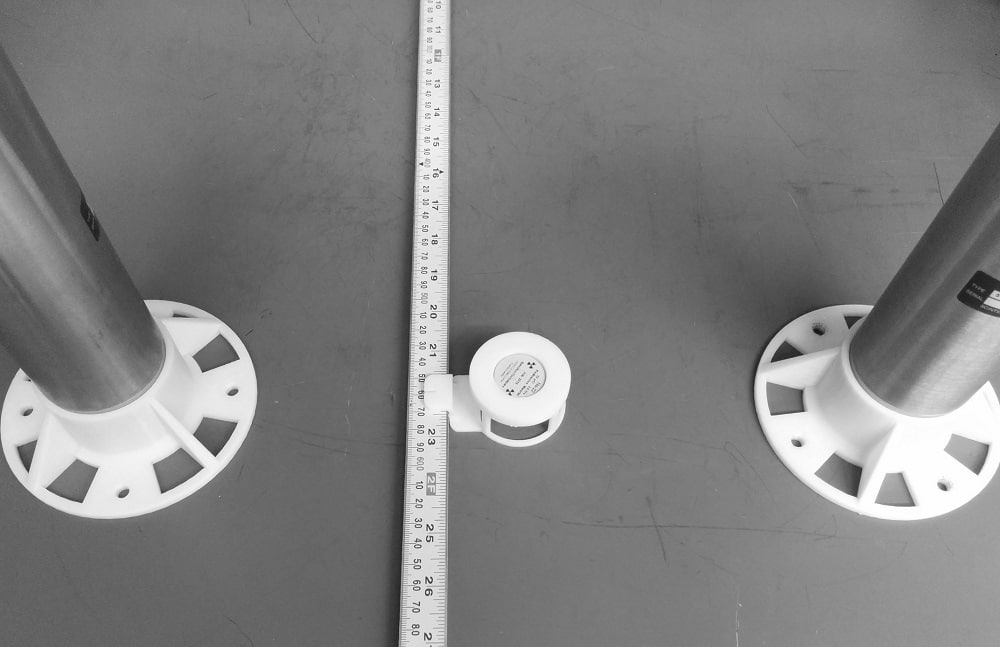}
\caption{Experimental setup for the PET scanning demonstration.  A source located between two detectors is moved along the shown ruler while event rates are measured.}
\label{petscansetup}
\end{figure}

\begin{figure}[h]
\centering
\includegraphics[width=3.4in]{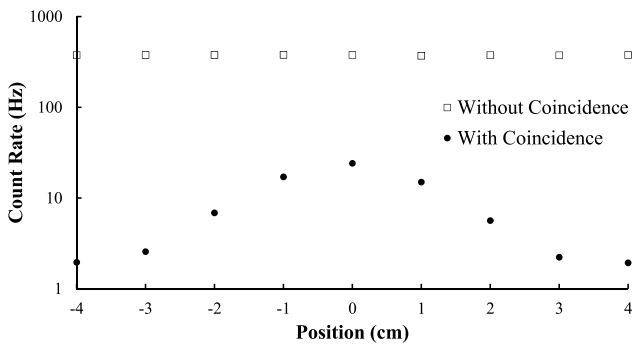}
\caption{PET scanning data showing variation of data rates with source position.  The use of coincidence detection clearly enhances the ability to infer the position of radiation as shown by the enhanced peak in data rates.}
\label{petscanplot}
\end{figure}

To further explore the nature of the 511 keV $\gamma$-rays emitted from positron annihilation, students are asked to explore how the nature of these $\gamma$-rays is used in PET scanning.  Two detectors are placed 50 cm apart with a \textsuperscript{22}Na source between them, as shown in Fig.~\ref{petscansetup}.  The source is then moved off-axis from between the detectors and the rate is tracked for each detector individually as well as for the coincidence between them. Figure~\ref{petscanplot} shows an example of the type of data acquired and the enhancement for locating the source of the radiation using the coincidence. 

\subsection{Low-Noise Compton Scattering}

\begin{figure}[h!]
\centering
\includegraphics[width=3.4in]{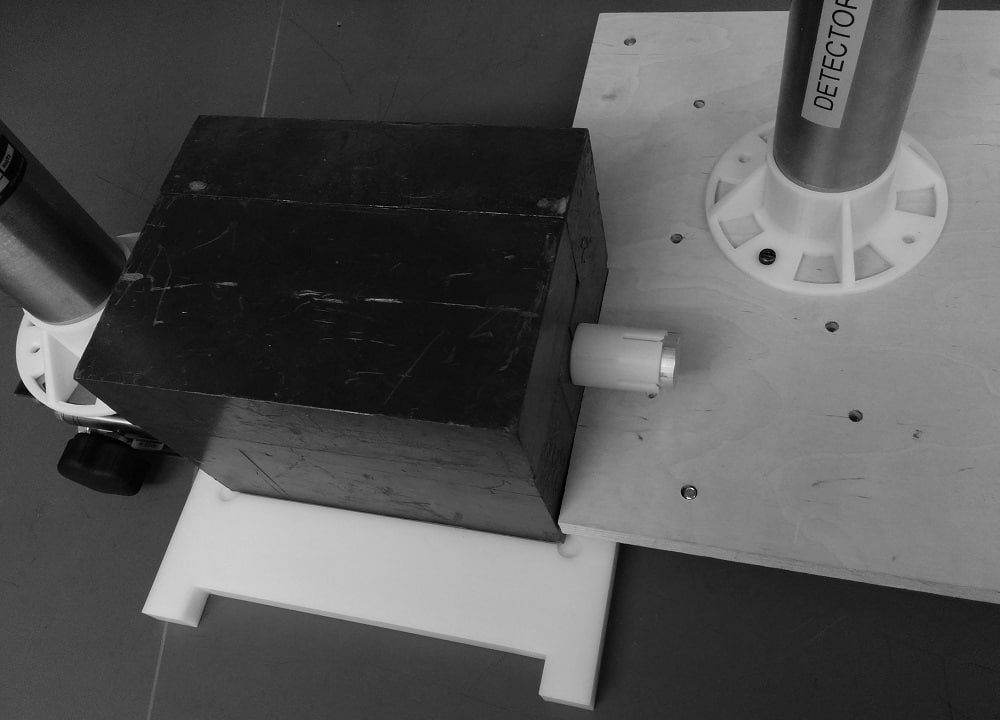}
\caption{Compton scattering setup to measure scattered $\gamma$-ray energy at a scattering angle of 60$^{\circ}$.}
\label{comptonsetup}
\end{figure}

This experiment utilizes the lead house for producing collimated 511 keV $\gamma$-ray pairs as shown in Fig.~\ref{comptonsetup}.    A 25 mm by 25 mm Aluminum cylinder is placed at one output port of the lead house to Compton scatter $\gamma$-rays.  One NaI detector is placed at the other port in order to detect the non-scattered $\gamma$-ray from the pair.  The second NaI detector is then placed to detect a scattered $\gamma$-ray at scattering angles of 0$^{\circ}$, 30$^{\circ}$, 60$^{\circ}$, and 90$^{\circ}$. 

\begin{figure}[h!]
\centering
\includegraphics[width=3.4in]{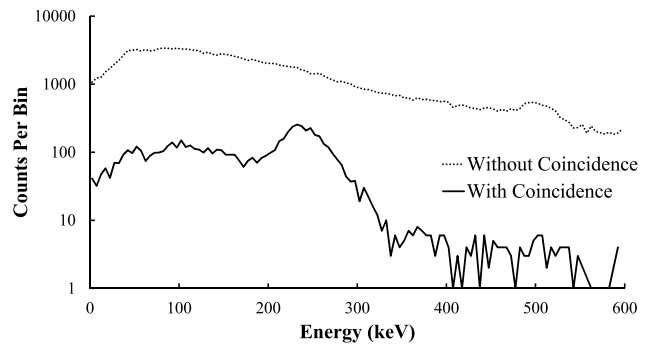}
\caption{$\gamma$-ray energy spectrum for Compton scattering at 90$^{\circ}$ with and without demanding a coincidence with the 511 keV pair $\gamma$-ray.  At this angle, the scattered photon energy is predicted to be 255 keV.  The peak at this energy is clearly enhanced using the coincidence technique.}
\label{comptoncoincidence}
\end{figure}

Figure~\ref{comptoncoincidence} shows the advantage of demanding a coincidence between both detectors in reducing the background noise in the signal.  This reduction in noise allows the students to make accurate measurements of energy for the scattered $\gamma$-rays that they can use to confirm Compton's relativistic prediction,

\begin{equation}
\label{compton}
\frac1{E'}-\frac1{E}=\frac1{E_0}(1-\cos(\theta)).
\end{equation}

\begin{figure}[h!]
\centering
\includegraphics[width=3.4in]{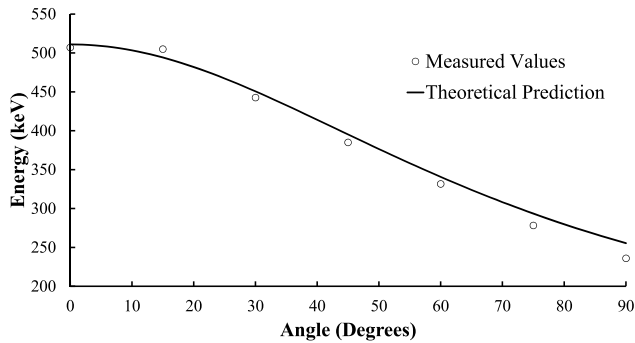}
\caption{Measured values for Compton scattered $\gamma$-rays compared to theoretical predictions.}
\label{comptonbyangle}
\end{figure}

Here, $E$ is the incident photon energy, $E'$ is the photon energy after scattering, $E_0$ is the electron rest energy, and $\theta$ is the scattering angle. Example measurements are shown in Fig.~\ref{comptonbyangle}.

\subsection{$\gamma$-Ray Velocity Measurement}

\begin{figure}[h!]
\centering
\includegraphics[width=3.4in]{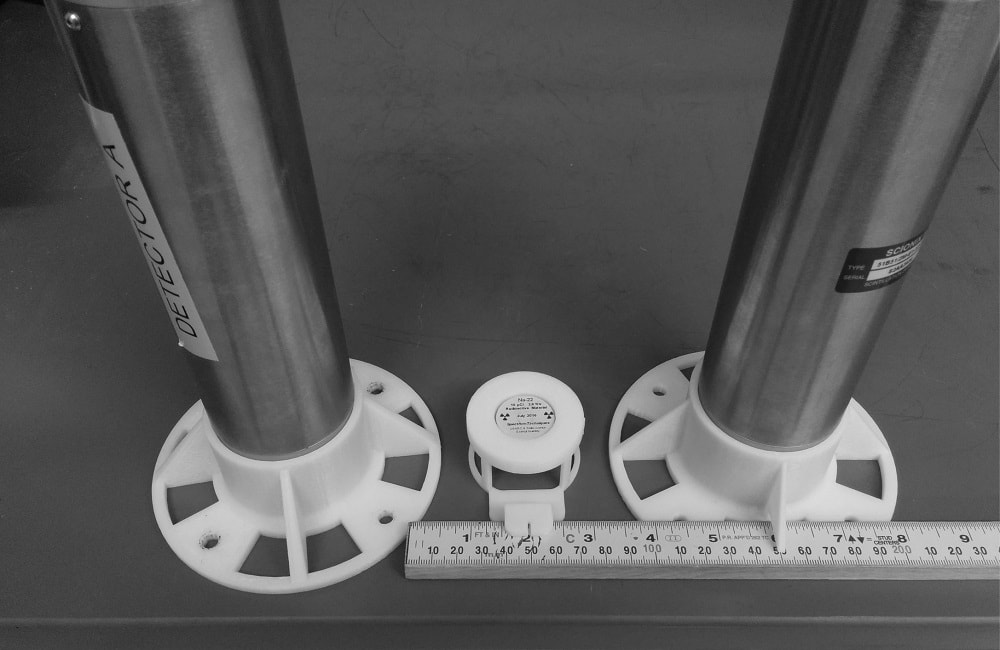}
\caption{Experimental setup for measurement of $\gamma$-ray velocity.  The detector on the right is moved along the ruler to create time delays that can be measured.}
\label{speedoflight}
\end{figure}

\begin{figure}[h!]%
    \centering
    \subfloat[]{{\includegraphics[width=3.4in]{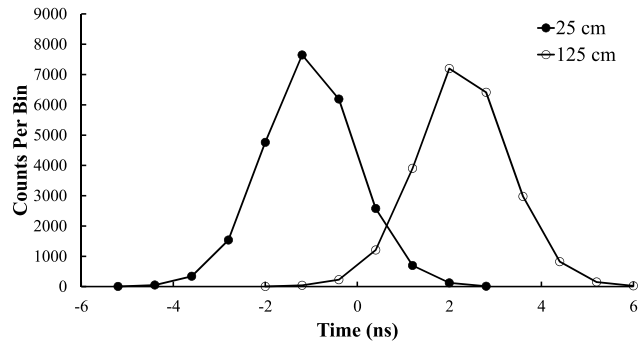} }}%
    \qquad%
    \subfloat[]{{\includegraphics[width=3.4in]{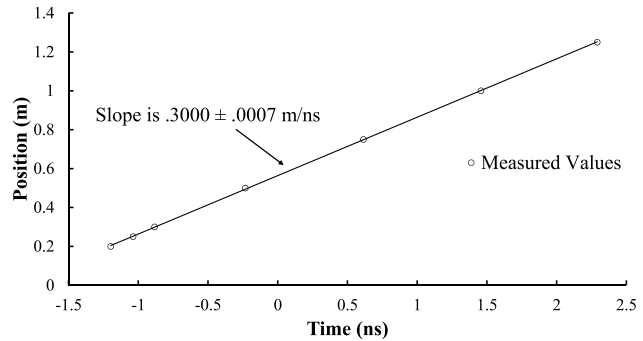} }}%
    \caption{(a) Timing histograms between two detectors shown for two different separation distances. These peaks can be fitted to determine their center for timing purposes. (b) Plot of the time delays with increasing distance. The slope represents the $\gamma$-ray velocity and is consistent with the speed of light as expected.}%
    \label{speedoflightdata}%
\end{figure}

In this experiment students make a direct measurement of the velocity of $\gamma$-rays.  This measurement helps to identify $\gamma$-rays as electromagnetic radiation by confirming that they indeed travel at the speed of light.  Figure~\ref{speedoflight} shows the experimental setup.  A \textsuperscript{22}Na source is set in between two detectors and the timing between signal arrivals is measured.  One of the detectors is then moved incrementally away from the source and the timing between signals is measured at each position.  Figure~\ref{speedoflightdata} shows an example of the data acquired and the evident time delays.

\section{Advanced Laboratory}
The advanced laboratory experiments are designed to allow students to delve further into experimental techniques using positron annihilation.  The individual experiments are more involved and offer the students the opportunity to explore on their own.

\subsection{Polarimetry of Annihilation Radiation}
\subsubsection{Experiment}
The back-to-back 511 keV $\gamma$-ray radiation from the positron-electron annihilation radiation provides an excellent opportunity to study an entangled pair system.  As discussed earlier in this paper, in solid materials the positron-electron pairs usually annihilate from the S=0 state, creating a pair of 511 keV $\gamma$-rays.  The spin state of the annihilating pair also determines the polarization of the entangled pair of $\gamma$-rays.  The linear polarization state of the $\gamma$-rays can be described\cite{17,18} as

\begin{equation}
\frac1{\sqrt{2}}\left(\ket{\updownarrow}_1 \ket{\leftrightarrow}_2 - \ket{\leftrightarrow}_1 \ket{\updownarrow}_2 \right).
\label{polarizationequation}
\end{equation}

Here, the $\updownarrow$ and $\leftrightarrow$ are polarizations and 1 and 2 are the back-to-back directions of travel.

To experimentally examine this polarization correlation, we can use the fact that Compton scattering is in fact angularly correlated to the polarization of the scattering $\gamma$-ray, a relationship first derived using QED by Klein and Nishina,\cite{28}

\begin{equation}
\label{KN}
\frac{d \sigma}{d \Omega} = \frac{r_e^2}{2}\left(\frac{E'}{E}\right)^2 \left[\frac{E'}{E}+\frac{E}{E'} -2 {\sin}^2(\theta){\cos}^2(\phi) \right].
\end{equation}

Here, $r_e$ is the classical electron radius (equal to 
$e^2 / (4\pi\epsilon_0 m_e c^2 )$
), $E$ is the energy of the incident photon, $E'$ is the energy of the scattered photon, $\theta$ is the scattering angle, and $\phi$ is the angle between the incident photon's polarization and the scattering direction.

\begin{figure}[h!]
\centering
\includegraphics[width=3.4in]{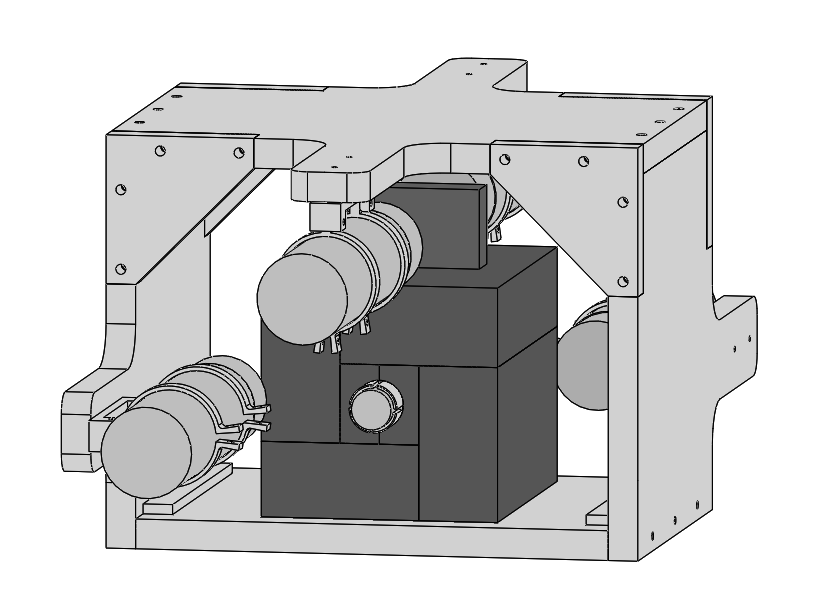}
\caption{Experimental setup for performing $\gamma$-ray polarimetry.  Aluminum scatterers are placed at both exits of the lead house.  NaI detectors are placed on both sides of the lead house to allow simultaneous detection of Compton scattering for both $\gamma$-rays in an annihilation 511 keV pair.}
\label{polarimetrysetup}
\end{figure}

To use this effect to perform polarimetry on the annihilation radiation, two aluminum cylinders are placed on either side of the lead house to Compton scatter $\gamma$-rays.  The 4 NaI detectors are placed around the lead house as shown in Fig.~\ref{polarimetrysetup} in order to detect $\gamma$-rays that scatter at 90$^{\circ}$.  The acquisition system is then set up to look for coincidences between pairs of detectors.  Because the pairs of $\gamma$-rays are in a state in which they are cross-polarized, coincidences should be favored in detectors that are separated axially by 90$^{\circ}$ ($\Delta \phi = 90^{\circ}$) over those separated by 0$^{\circ}$ ($\Delta \phi = 0^{\circ}$) and 180$^{\circ}$ ($\Delta \phi = 180^{\circ}$).  The relative probability for pairs of scattered $\gamma$-rays to be detected is expressed by

\begin{equation}
\rho = \frac{R_{90^{\circ}}}{R_{0^{\circ}}} = 1+\frac{2 {\sin}^4(\theta)}{\gamma^2-2 \gamma {\sin}^2(\theta)},
\label{enhancement}
\end{equation}

where

\begin{equation}
\gamma = 2-\cos(\theta)+\frac1{2-\cos(\theta)}.
\end{equation}

Here, $R_{90^{\circ}}$ and $R_{0^{\circ}}$ are the event rates at $\Delta \phi = 90^{\circ}$ and 0$^{\circ}$ respectively. It should be noted that $R_{0^{\circ}}=R_{180^{\circ}}$.
This follows from Eq.~(\ref{polarizationequation}) and Eq.~(\ref{KN}), a full derivation of which can be seen elsewhere.\cite{17,18}

Assuming infinitesimally small solid angles for the detectors, Eq.~(\ref{enhancement}) predicts $\rho=2.6$ when $\theta = 90^{\circ}$.  Experimentally we found $\rho=2.4\pm 0.1$.  This was looking at coincidence rates for events in the energy range 245--265 keV, a range that is centered on the predicted energy of Compton scattered annihilation gamma rays of 255 keV.  Given that the solid angles of acceptance are substantial this is a surprisingly good result. This indicates that although the acceptance angles are not ideal, the desired scattering angles are highly favored. 

\subsubsection{Historical Context}
In addition to examining the polarization of the $\gamma$-rays from positron annihilations, this experiment replicates some of the earliest experiments used to look at entanglement and the EPR paradox.\cite{19}  This annihilation pair was first theoretically examined independently in two papers.\cite{17,18} Experimentally, this system was then studied by Wu and Shaknov\cite{20} in 1950.  Later, as other theoretical works were attempting to understand the paradox of EPR,\cite{21} Bohm and Ahronov\cite{22,23} argued in 1957 that this earlier positron polarimetry work in fact was evidence that the properties of quantum mechanics criticized by the EPR paradox were in fact real properties of matter.

After the work of Bell\cite{24} in 1964, it was understood that similar correlation experiments could be used to rule out all hidden variable theories.  In light of this work, Kasday et al.\cite{25} performed an update to the experiment of Wu and Shaknov\cite{20} in 1974 in order to interpret the results of the technique as a Bell's Inequality experiment.  In that work and later work\cite{26} it has been shown that hidden variable theories can in fact replicate the result of the Klein-Nishina equation.  Thus, the result of any correlated polarization experiment that uses Compton scattering as the mechanism for the polarimetry can also be reproduced by local hidden variable theories.  Therefore, while these types of experiments are important for historical context, they have been replaced by other techniques in modern experiments.

\subsection{Positron Lifetime Spectroscopy}
When introduced into a solid material, a positron will follow one of a few paths to annihilation.  On the short time scale they can form para-Ps or find an electron to directly annihilate with.  Direct annihilation typically takes a few hundred ps while para-Ps decays exponentially with a lifetime of 125 ps.  Both of these time scales are significantly less than the 2.4 ns timing resolution of our detectors and thus will simply show up as a timing peak in our system.

The other path for annihilation occurs when the positron binds with an electron and forms ortho-Ps.  This three-$\gamma$ annihilation has a much longer lifetime of 142 ns in vacuum.  In a solid material this will be reduced greatly by the interaction of the Ps with the material.  The positron in the bound system can find an electron other than its bound partner to annihilate with, reducing its lifetime in a process called ``pick-off''.

Thus, the lifetime of o-Ps in a material can be used to study the void properties of the materials it is in.  When voids of {\raise.17ex\hbox{$\scriptstyle\sim$}}1 nm in size are available, the lifetime can be used to quantitatively measure the size of the voids using the Tao-Eldrup model,\cite{27}

\begin{equation}
\tau = \frac{0.5~\textrm{ns}}{1-\frac{R}{R+0.166~\textrm{nm}}+\frac1{2\pi}\sin{\left(\frac{2\pi R}{R+0.166~\textrm{nm}}\right)}}.
\label{TEeq}
\end{equation}

This semi-empirical equation gives the lifetime of ortho-Ps ($\tau$) as a function of the radius of the voids in the material (R).

\begin{figure}[h!]
\centering
\includegraphics[width=3.4in]{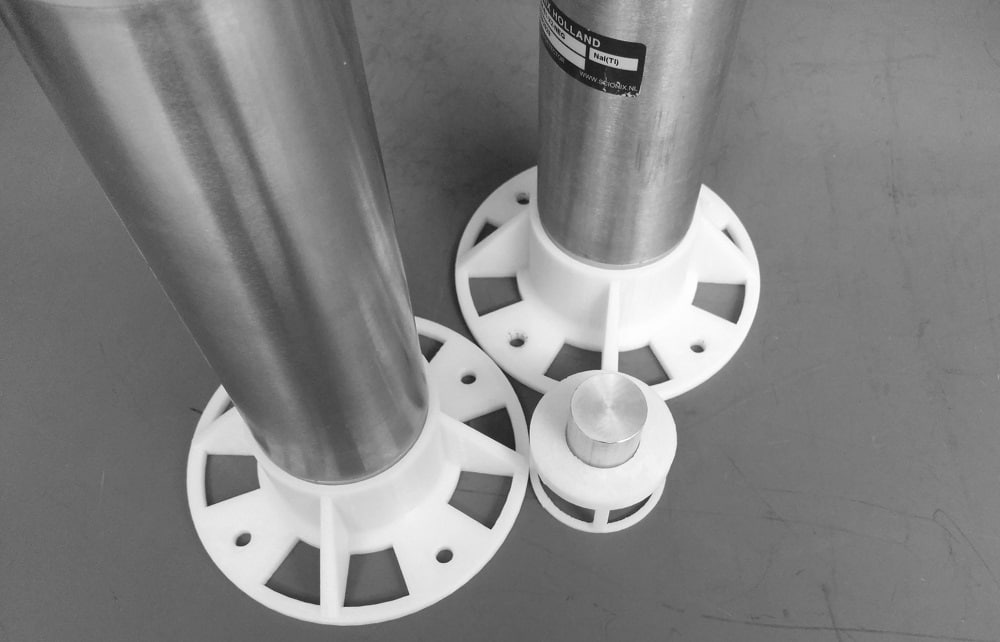}
\caption{Positron lifetime spectroscopy experiment setup.}
\label{lifetimesetup}
\end{figure}

\begin{figure}[h!]
\centering
\includegraphics[width=3.4in]{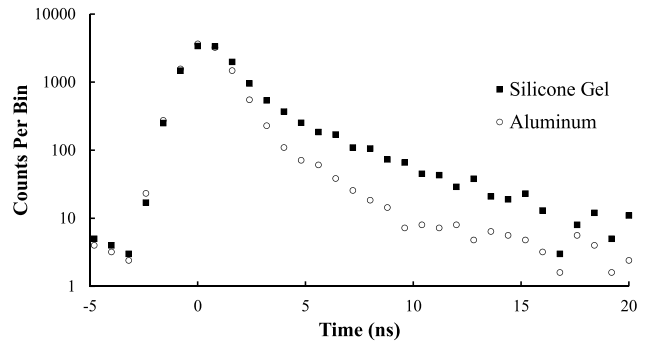}
\caption{Positron lifetime spectroscopy data. Silicone and Aluminum show clear differences in positron/positronium lifetime due to the difference in positron interactions in the two materials.}
\label{lifetimeplot}
\end{figure}

The setup for the experiment is simple.  A \textsuperscript{22}Na source (the 2 $\mu$Curie source with a thin Mylar window) is placed at 90$^{\circ}$ to two detectors as shown in Fig.~\ref{lifetimesetup}.  A sample material of interest is placed on top of the source and the timing is measured between 1275 keV $\gamma$-rays and annihilation $\gamma$-rays.  Figure~\ref{lifetimeplot} shows the timing spectra of two samples.  The Aluminum, as a metal, does not form Ps and thus has a shorter lifetime than the Silicone sample.  The Silicone, with a Ps lifetime of about 4 ns can be studied quantitatively by students if desired.

\subsection{3-$\gamma$ Annihilation Radiation}
The o-Ps 3-$\gamma$ decay channel is difficult to observe without concerted effort.  In a solid material, the dominant effect is pick-off and 2-$\gamma$ decay.  In order to achieve a significant amount of 3-$\gamma$ annihilation, the Ps needs to form in a diffuse environment such as a gas.

We achieve this environment by placing fumed silica powder on a \textsuperscript{22}Na source.  The powder helps keep the positrons from the \textsuperscript{22}Na source contained to a small space, but also allows for any Ps formed to exist in large open volumes.  In our system, the air that fills the space within the powder is still able to reduce the o-Ps lifetime from 142 ns to {\raise.17ex\hbox{$\scriptstyle\sim$}}70 ns via pick-off.  Nevertheless, a large fraction of the o-Ps can decay via 3 $\gamma$-rays.

\begin{figure}[h!]
\centering
\includegraphics[width=3.4in]{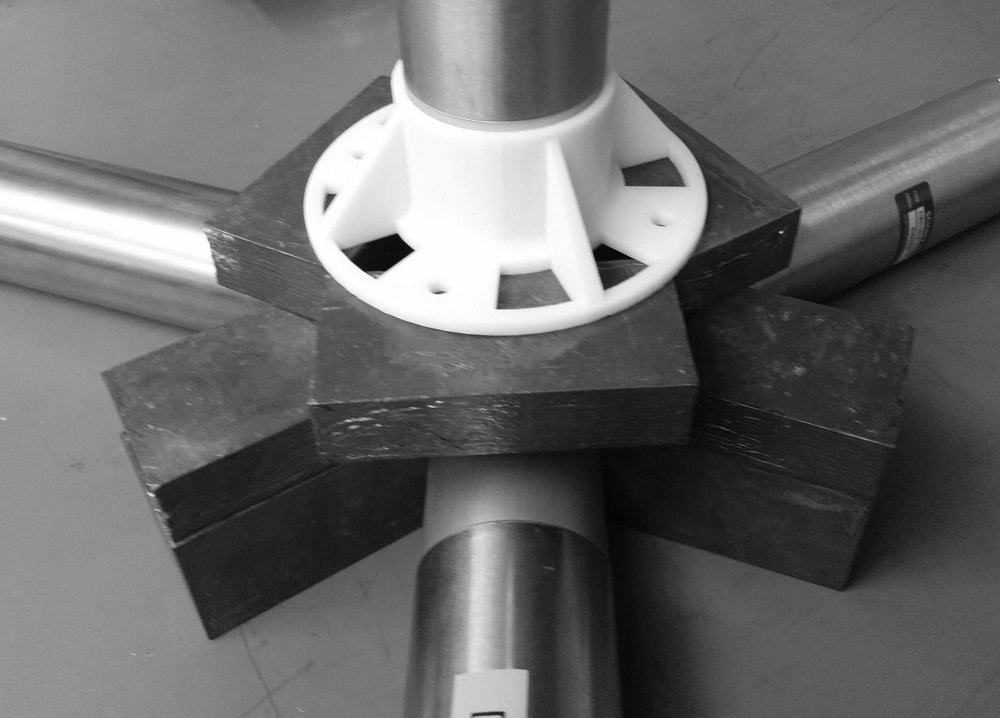}
\caption{Experimental setup for 3-$\gamma$ detection.   Three detectors are placed so as to be coplanar and separated by 120$^{\circ}$.  Lead is placed between the detectors to reduce scattering between them.  The top detector is used to detect the coincident 1275 keV $\gamma$ to further reduce accidental coincidences.}
\label{threegammasetup}
\end{figure}

The experiment aims to confirm the existence of the 3-$\gamma$ decay channel.  To accomplish this we use four detectors as shown in Fig.~\ref{threegammasetup}.  Three of these detectors are co-planar while the fourth detector is placed above the others.  Coincidence is demanded between all 4 detectors.  Events that are detected are filtered in two ways.

\begin{enumerate}
\item The three co-planar detector events are examined for timing to ensure that the detections occurred with a 10 ns window.
\item The fourth detector is required to have an energy of 1275 keV.
\end{enumerate}

\begin{figure}[h!]
\centering
\includegraphics[width=3.4in]{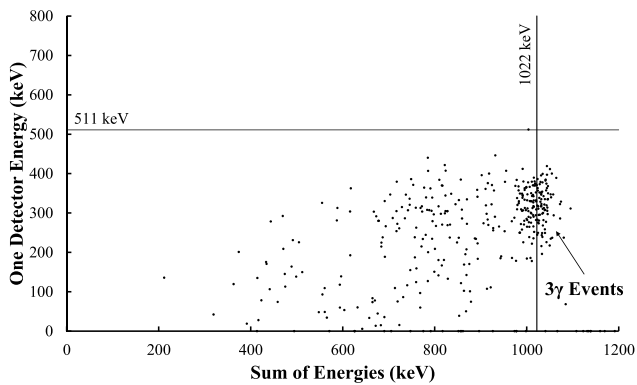}
\caption{Scatter plot of detected events for the 3-$\gamma$ detection experiment.  Lines at 511 keV for the single detector energy and 1022 keV for the 3 detector energy sum are shown for reference.  A clear accumulation of events can be seen in the expected region for 3-$\gamma$ events as indicated.}
\label{threegammadata}
\end{figure}

The resulting data can be plotted as shown in Fig.~\ref{threegammadata}.  This figure shows the energy of a single detector vs. the sum of the energies of the three co-planar detectors.  A clear signal of 3-$\gamma$ events is the highlighted area in the figure.  The highlighted area is at 1022 keV in the sum of the detectors indicating that all three  detectors add up to the appropriate energy of the positron-electron annihilation.  The area does not extend up to 511 keV for a single detector.  This is important as it indicates that the signals are not 2-$\gamma$ pairs in which one $\gamma$-ray scattered from one detector into the other.

\section{Conclusion}

The experiments presented here represent a diverse and flexible set of learning experiments for students.  In addition to the examples presented here, there are opportunities for students to explore further.  Examples include.

\begin{enumerate}
\item Looking at a wide variety of samples to develop models of the Ps lifetime in materials.
\item Replacing the air in the fumed silica with other gasses to get better measurements of the o-Ps lifetime.
\item Increasing the precision of the speed of light measurement with more data acquisition and precise position measurements.
\end{enumerate}

In the future we hope to develop the apparatus further to allow for more flexibility.  Ideas include:

\begin{enumerate}
\item Placing a source in a container in which gas content and pressure can be controlled to allow for careful measurement of the o-Ps lifetime as a test of QED.\cite{12}
\item Passing the detector signals through an FPGA in order to get accurate rate measurements when rates are higher than accessible by the oscilloscope.
\item Expanding the software to include more data presentation tools.
\end{enumerate}

Finally, in order to promote the adoption of these experiments at other institutions we have made available all the work presented here on the St.\@ Olaf Physics Department website.  This includes detailed drawings of the apparatus, software, and lab manuals. These materials will be updated with future progress.

\begin{acknowledgments}
We would like to thank St.\@ Olaf College for their financial support in this work as well as our colleagues in the St.\@ Olaf Physics department who have provided valuable feedback in the development of these experiments.
\end{acknowledgments}

\end{document}